\documentstyle[11pt,paspconf,epsf]{article}

\newcommand{\be}{\begin{equation}}
\newcommand{\ee}{\end{equation}}
\newcommand{\sm}{\mbox{M}_{\odot}}

\begin{document}

\title{Searching for MACHOs in Andromeda with INT}
\author{Eamonn Kerins}
\affil{Theoretical Physics, 1 Keble Road, Oxford
OX1 3NP, UK}

\begin{abstract} 
Two teams are using the Isaac Newton Telescope to conduct a
microlensing search for massive compact halo objects (MACHOs) in
Andromeda. We discuss both the motivation of the surveys and the
obstacles they must overcome. The key to success is the spatial
distribution of detected events. We present a detailed simulation of
this observable.
\end{abstract}


\section{Microlensing towards Andromeda}

Two experiments, AGAPE (\cite{ansari}) and MEGA (\cite{crotts}), have
recently commenced observations on the Isaac Newton Telescope
Wide-field Camera (INT WFC) with the hope of detecting, for the first
time, massive compact halo objects (MACHOs) in another
galaxy---Andromeda (M31).
\begin{figure}[h]
\begin{minipage}{6.cm}
\plotfiddle{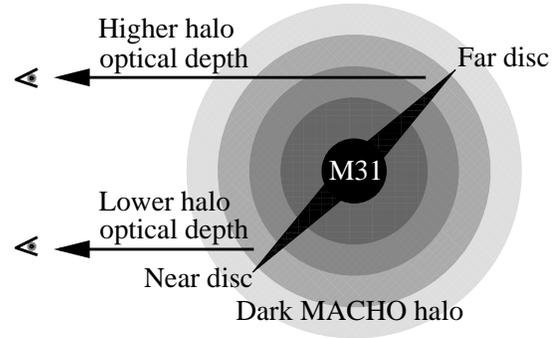}{0cm}{270}{100}{100}{80}{0}
\caption{Near-far microlensing asymmetry towards Andromeda. The
optical depth is larger towards the far disc than towards the
near side if MACHOs are present in Andromeda's halo.  The effect is
less pronounced if its halo is flattened.}
\label{f1}
\end{minipage}
\end{figure}
Whilst the claim of MACHO detection in our own Galaxy remains hotly
disputed, Andromeda's $77^\circ$ inclination provides a signature
which {\em unambiguously}\/ betrays the presence of MACHOs: near-far
asymmetry (Fig.~\ref{f1}). Such spatial asymmetry is unique to M31
MACHOs: it does not occur for foreground Milky Way lenses or for
stellar lenses or variable stars in the M31 disc.  M31 has other
attributes that make it ideal for microlensing studies. It provides
${\cal O}(10^3)$ times as many sources than available towards the
Magellanic clouds or Milky Way bulge. Its nearby location (770~kpc)
permits detailed study of its surface brightness and rotation curve
profiles, thus reducing the mass modeling uncertainties which hinder
the interpretation of Milky Way microlensing events.

However, the major problem facing surveys probing
beyond our own Galaxy is that background stellar fields are not
resolved. This means that if a microlensing event is detected the
source star cannot be identified. It also means that the
microlensing signal is contaminated by the background flux of the
other stars, making it harder to detect. The problem is made
worse by changes in detector alignment, sky backgrounds and observing
conditions (especially seeing variations) from one observation to the
next. To overcome this AGAPE has developed the {\em superpixel}\/
technique, whilst MEGA employs {\em difference imaging}\/. Both
techniques minimize detector, sky and seeing temporal variations so as
to maximize the microlensing signal. The two techniques have both been proven
in pilot studies and their sensitivity is practically photon noise limited.

\section{INT simulations and predictions}

\begin{figure}
\plotone{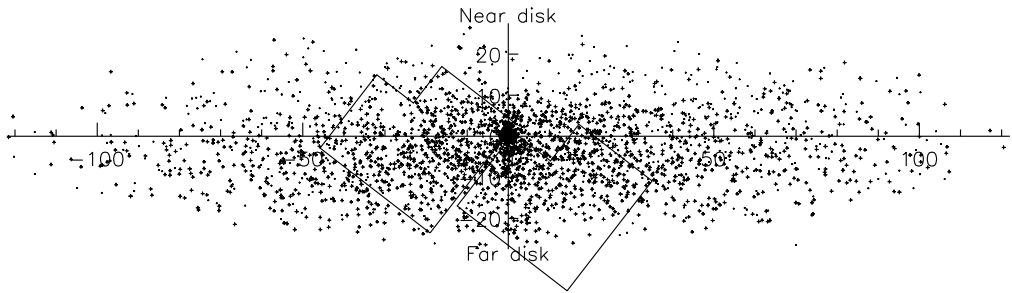}
\caption{Simulated spatial distribution for 4000 ``detected'' events.}
\label{f2}
\end{figure}
We have implemented detailed simulations for the superpixel
technique. Fig.~\ref{f2} shows the predicted spatial distribution for
4000 ``detected'' events (equivalent to about $20$ seasons of data) with
the axes in arcmins. It assumes 20~min $V$-band exposures and allows
for the effects of variable sky/moon backgrounds, instrument down-time
and typical weather interruptions, all of which determine the
experimental flux and temporal sensitivity and thus the observed
spatial distribution. The sky background in particular determines the
spatial cutoff beyond which no events are detected. The M31 halo
events (larger dots) and Milky Way halo and M31 bulge/disc events
(smaller dots) are shown for spherical haloes of $0.3~\sm$ lenses and
bulge/disc components with a solar-neighbourhood mass function. The
locations of the two AGAPE/MEGA INT WFC fields are also indicated. The
near-far asymmetry in the M31 halo lenses is clearly evident
(e.g. compare event densities at $\pm 15$ arcmin on the minor
axis). The simulations predict up to $\sim 50$ events per INT WFC
field per season if MACHOs in the range $\sim10^{-3} - 3~\sm$ comprise
the dark matter. This is sufficient to establish near-far asymmetry
within a few INT observing seasons or, if asymmetry is absent, to
begin placing constraints on the M31 MACHO contribution.

\end{document}